\documentclass[aps,superscriptaddress,amsmath,amssymb,twocolumn,showpacs,floatfix,reprint]{revtex4-1}
\usepackage{bm}
\usepackage{graphicx}
\usepackage[colorlinks=true, urlcolor=blue, linkcolor=blue, citecolor=blue, pdftex]{hyperref}
\usepackage{multirow}
\usepackage{float}

\begin{document}

\title{Transformer variational wave functions for frustrated quantum spin systems}

\author{Luciano Loris Viteritti}
\thanks{These authors contributed equally.}
\affiliation{Dipartimento di Fisica, Universit\`a di Trieste, Strada Costiera 11, I-34151 Trieste, Italy}
\author{Riccardo Rende}
\thanks{These authors contributed equally.}
\affiliation{International School for Advanced Studies (SISSA), Via Bonomea 265, I-34136 Trieste, Italy}
\author{Federico Becca}
\affiliation{Dipartimento di Fisica, Universit\`a di Trieste, Strada Costiera 11, I-34151 Trieste, Italy}

\date{\today}

\begin{abstract}
The Transformer architecture has become the state-of-art model for natural language processing tasks and, more recently, also for computer 
vision tasks, thus defining the Vision Transformer (ViT) architecture. The key feature is the ability to describe long-range correlations 
among the elements of the input sequences, through the so-called self-attention mechanism. Here, we propose an adaptation of the ViT architecture 
with complex parameters to define a new class of variational neural-network states for quantum many-body systems, the ViT wave function. We apply this idea to the one-dimensional $J_1$-$J_2$ Heisenberg model, demonstrating that a relatively simple parametrization gets excellent results 
for both gapped and gapless phases. In this case, excellent accuracies are obtained by a relatively shallow architecture, with a single layer of 
self-attention, thus largely simplifying the original architecture. Still, the optimization of a deeper structure is possible and can be used for more challenging models, most notably highly-frustrated systems in two dimensions. The success of the ViT wave function relies on mixing 
both local and global operations, thus enabling the study of large systems with high accuracy.
\end{abstract}

\maketitle

{\it Introduction.} 
Variational approaches for studying quantum many-body systems have proved fundamental for understanding the properties of extremely complicated 
physical systems, famous examples being the Bardeen-Cooper-Schrieffer state~\cite{bardeen1957} and Laughlin~\cite{laughlin1983} wave functions 
to explain superconductivity and fractional quantum Hall effect, respectively. Given the exponential growth of the many-body Hilbert space, a 
compact representation of the ground state, encoding the correct physical properties, is a highly non-trivial task for strongly-interacting
systems. Recently, a class of wave functions, based on neural networks, has been introduced and developed~\cite{carleo2017,glasser2018}. Starting 
from Restricted Boltzmann Machines (RBMs)~\cite{carleo2017}, which are the simplest neural-network {\it Ansatz} (namely only one fully-connected 
hidden layer), numerous studies have been carried out testing different types of architectures; examples include Convolutional-Neural Networks 
(CNNs)~\cite{choo2019,liang2018,chen2022,roth2021}, Recurrent-Neural Networks (RNNs)~\cite{hibat2020,hibat2022}, Autoregressive-Neural 
Networks~\cite{luo2021,sharir2020}, but also combinations of neural networks with standard variational wave functions (e.g., Gutzwiller-projected 
fermionic ones)~\cite{nomura2017,ferrari2019}.

In the last few years, the Transformer architecture~\cite{vaswani2017} has become the state-of-art choice in natural-language processing tasks. 
Its key feature is the ability to model relationships among all elements of an input sequence (regardless of their positions), by efficiently 
\textit{transforming} input sequences into abstract representations. Inspired by successes in natural-language processing, very small modifications 
led to the so-called Vision Transformer (ViT)~\cite{dosovitskiy2021}, which has been applied to image classification tasks, achieving competitive 
results with respect to state-of-art deep CNNs, while being much more efficient than them. Within many-body problems, Transformer networks have 
recently been employed in the context of lattice gauge theories~\cite{luo2021}, to perform quantum tomography in presence of noise~\cite{cha2021},
and for real- and imaginary-time evolutions of quantum systems~\cite{luo2022}.

In this Letter, we demonstrate that the ViT architecture can be adapted to define a new class of neural-network quantum states, here dubbed as ViT 
wave functions. We apply our {\it Ansatz} to the one-dimensional $J_1$-$J_2$ Heisenberg model, whose Hamiltonian is defined by 
\begin{equation}\label{eq:J1J2_ham}
\hat{H} = J_1\sum_{R} \hat{\boldsymbol{S}}_{R}\cdot\hat{\boldsymbol{S}}_{R+1} + J_2\sum_{R} \hat{\boldsymbol{S}}_{R}\cdot\hat{\boldsymbol{S}}_{R+2} \,
\end{equation}
where $\hat{\boldsymbol{S}}_{R} = (S_R^x, S_R^y, S_R^z)$ is the $S = 1/2$ spin operator at site $R$ and $J_1>0$ and $J_2 \ge 0$ are nearest- and 
next-nearest-neighbor antiferromagnetic couplings, respectively. Its phase diagram is well established by analytical 
and numerical studies~\cite{white1996}. For small values of $J_2/J_1$, the ground state has power-law spin-spin correlations and the excitation 
spectrum is gapless; for large values of $J_2/J_1$, the ground state is two-fold degenerate, leading 
to long-range dimer order (but exponentially decaying spin-spin correlations), and the spectrum is fully gapped. These two phases are separated by 
a critical point at $(J_2/J_1)_c = 0.241167\pm 0.000005$~\cite{eggert1996,sandvik2010}. Interestingly, for $J_2/J_1>0.5$, incommensurate (but 
short-range) spin-spin correlations have been found, whereas dimer–dimer correlations are always commensurate. In the following, we assess the
ground-state properties of the $J_1$-$J_2$ model on finite clusters, imposing periodic boundary conditions.

From the numerical perspective, density-matrix renormalization group (DMRG)~\cite{white1992} or its modern variations based upon tensor networks 
{\it Ans\"atze}~\cite{schollwock2011} represent one of the few approaches that can accurately assess the ground-state properties of frustrated
systems in one dimension, as the $J_1$-$J_2$ model of Eq.~\eqref{eq:J1J2_ham}. In fact, the main limitation to the use of quantum Monte Carlo 
techniques~\cite{becca2017} relies on the unknown sign structure of the ground-state wave function, which prevents one to perform unbiased 
projection techniques (except for $J_2=0$, where the so-called Marshall sign rule applies~\cite{marshall1955}). The non-trivial sign structure 
represents also an obstacle to the definition of accurate variational wave functions. For example, Gutzwiller-projected fermionic 
states~\cite{ferrari2018} have a limited power to reproduce the correct signs of the ground state for $J_2/J_1>0.5$~\cite{viteritti2022}. 
By contrast, RBM states are able to reach an excellent accuracy; however, they suffer from poor scaling behavior, due to their {\it fully-connected}
structure in which a single hidden layer is connected to all physical degrees of freedom~\cite{viteritti2022}. This fact limits the applicability 
of RBMs to relatively small clusters. In this respect, CNN wave functions have been introduced to deal with {\it local} structures and deep 
architectures are necessary to build long-range correlations, thus introducing severe problems in the optimization procedure (e.g., diverging or 
vanishing gradients). RNNs {\it Ans\"atze} have been also considered, which recurrently process inputs of a sequence one by one, implying that 
they cannot be parallelized; in addition, since not all elements of the network are directly connected, long-range correlations are built from 
short-range ones, thus making the learning process not straightforward~\cite{bengioRNN1994}.

In order to overcome these problems, we propose a simplified version of the standard ViT architecture. The main advantage of this {\it Ansatz}
lies in the possibility to mix both local and global structures, thus limiting the number of variational parameters and simplifying the learning 
process (see below). We emphasize that a complex parametrization is adopted without an {\it a priori} encoding of the sign structure (i.e., no 
information about the exact signs). In this work, we show that the ViT wave function can reach very high accurate results compared to DMRG 
calculations, even on large clusters, with less then one thousand parameters and few computational resources compared to other neural-network 
wave functions. Most importantly, the ViT accuracy can be systematically improved by changing the hyper-parameters of the architecture.

%%%%%%%%%%%%%%%%%%%%%%%%%%%%%%%%%%%%%%%%%%%%%%%%%%%%%%%%%%%%%%%%%%%%%%%%%%%%%%%%%%%%%%%%%%%%%%%%%%%%%%%%%%%%%%%%%%%%%%%%%%%%%%%%%
\begin{figure}
\includegraphics[scale = 0.4]{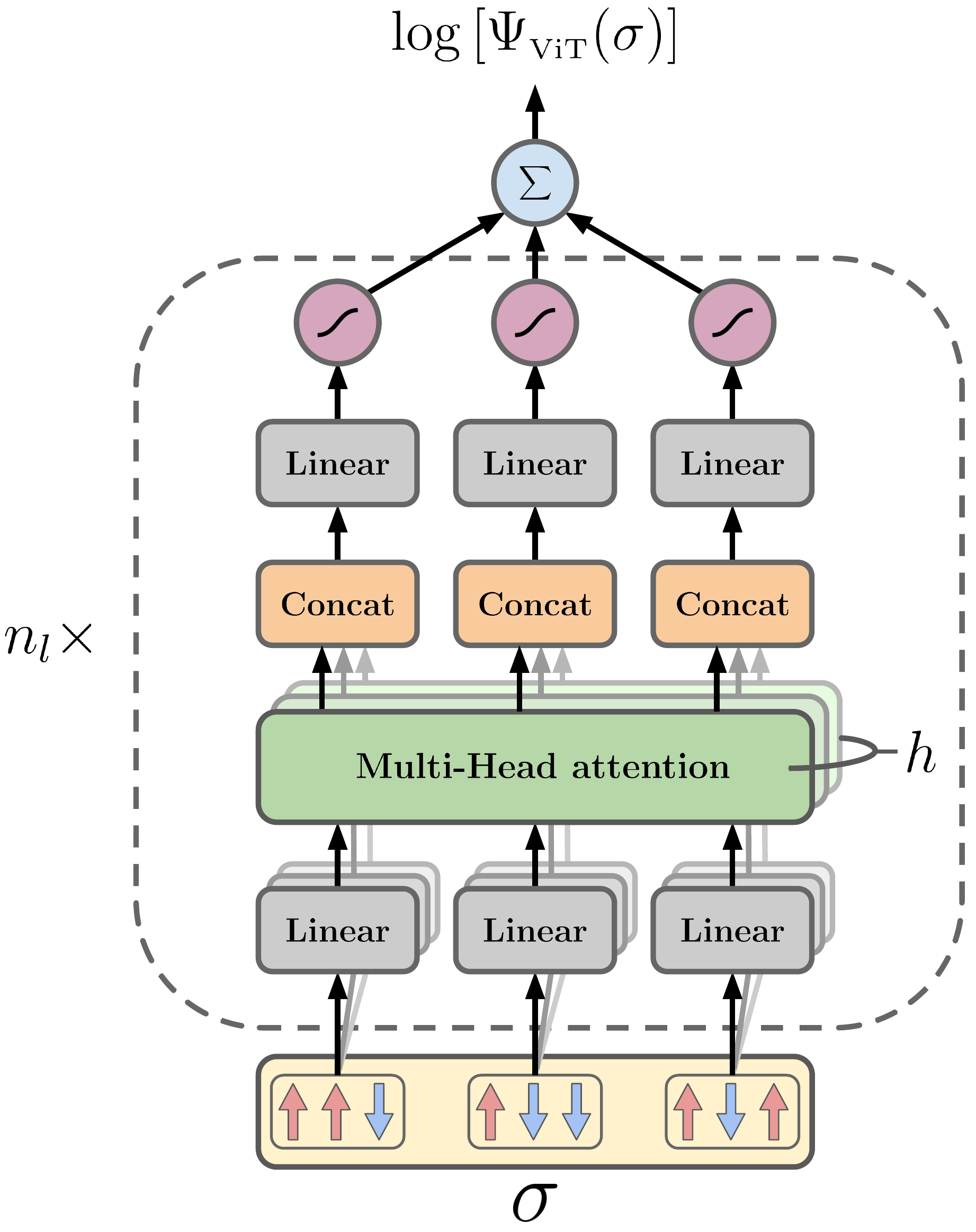}
\caption{\label{fig:transf}
Scheme of the ViT wave function. The input spin configuration $\sigma$ is split into patches of size $b$ (which define a set of $N$ vectors of
dimension $b$). Each of them is linearly projected $h$ times with different linear projections to produce $N$ vectors of dimensions $r=d/h$. 
Then the attention function is applied in parallel and the $h$ different $r$ dimensional output vectors $\boldsymbol{A}_i^{\mu}$ are obtained. 
Then, they are concatenated to a $d$ dimensional vector $\text{Concat}(\boldsymbol{A}_i^{1},\dots,\boldsymbol{A}_i^{h})$ and, after another 
linear projection, the non-linear function log[cosh($\cdot$)] is applied. This architecture can be replicated and stacked $n_l$ times. The last 
layer simply sums all the outputs and returns the logarithm of the ViT wave function.} 
\end{figure}

\begin{figure}
\includegraphics[width=\columnwidth]{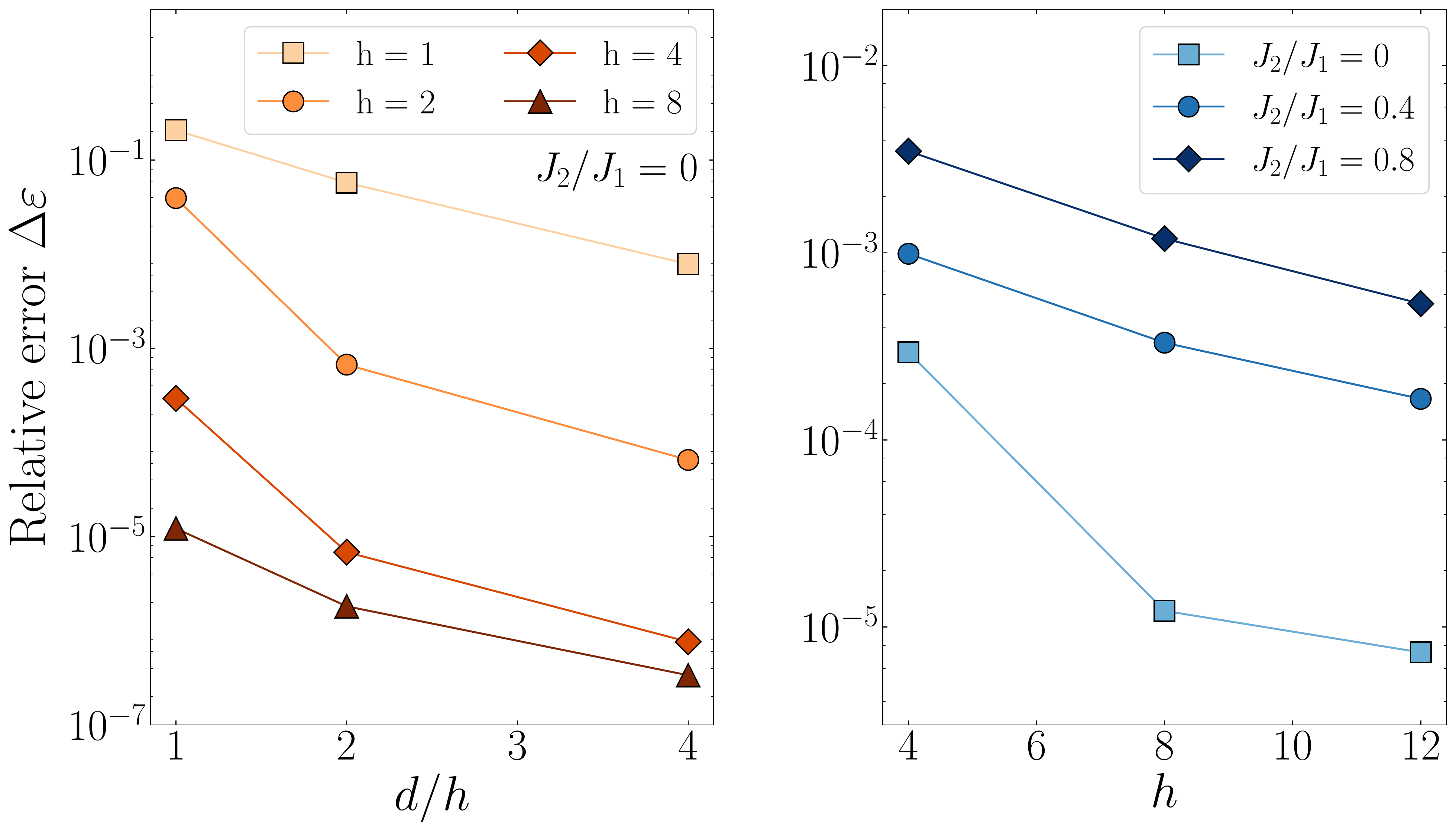}
\caption{\label{fig:acc}
Relative error $\Delta \varepsilon =|(\text{E}_{\rm ViT} - \text{E}_{\rm DMRG})/\text{E}_{\rm DMRG}|$ of the ViT wave function by varying the 
hyper-parameters of the architecture for a cluster with $L=100$ sites. Left panel: $\Delta \varepsilon$ as a function of $d/h$, with a fixed number 
of heads $h$, for the unfrustrated case. Right panel: $\Delta \varepsilon$ as a function of the number of heads $h$, with $d/h=1$, for different 
values of frustration ratio. The reference energies are computed by DMRG~\cite{pippan2010} with a bond dimension up to $\chi=600$ obtaining 
$E/J_1=-0.4432295$ for $J_2/J_1=0$, $E/J_1=-0.3803882$ for $J_2/J_1=0.4$, and $E/J_1=-0.4216664$ for $J_2/J_1=0.8$.}
\end{figure}
%%%%%%%%%%%%%%%%%%%%%%%%%%%%%%%%%%%%%%%%%%%%%%%%%%%%%%%%%%%%%%%%%%%%%%%%%%%%%%%%%%%%%%%%%%%%%%%%%%%%%%%%%%%%%%%%%%%%%%%%%%%%%%%%%

{\it Methods.}
The fundamental ingredient of a Transformer is the {\it self-attention mechanism}. Given a sequence of $N$ input vectors 
$(\boldsymbol{x}_1, \dots, \boldsymbol{x}_N)$, for each of them three new vectors are computed, $\boldsymbol{q}_i = \text{Q} \boldsymbol{x}_i$, 
$\boldsymbol{k}_i = \text{K}\boldsymbol{x}_i$, and $\boldsymbol{v}_i = \text{V}\boldsymbol{x}_i$, where $\text{Q}$, $\text{K}$, $\text{V}$ are 
generic rectangular matrices of parameters. The attention vectors are then constructed, $\boldsymbol{A}_i = \sum_{j = 1}^N 
\alpha(\boldsymbol{q}_i, \boldsymbol{k}_j) \boldsymbol{v}_j$, where the attention weights $\alpha(\boldsymbol{q}_i, \boldsymbol{k}_j)$ determine 
how much the $j$-th input vector should contribute to $\boldsymbol{A}_i$, which is the subsequent representation of the $i$-th input. The functional
form of these weights can be chosen according to the task~\cite{taysurvey2020}. To improve the performance of the model, multi-head attention 
can be considered, where a set of matrices $Q^\mu$, $K^{\mu}$, and $V^{\mu}$, with $\mu=1,\dots, h$ (with $h$ the {\it number of heads}) 
is defined, thus leading to a set of attention vectors $\boldsymbol{A}^{\mu}_i$. The latter ones are computed in parallel, concatenated together, 
and linearly combined. Finally, each output vector of the multi-head attention is fed separately and identically to a non linearity. In general, 
this whole architecture is replicated $n_l$ times.

Our goal is to use the Transformer to parameterize the many-body wave function, in order to map spin configurations of the Hilbert space 
$\sigma = (\sigma_1, \dots, \sigma_L)$, with $\sigma_R = 2 S_R^z = \pm 1$, to complex numbers $\Psi(\sigma)$. We take inspiration from the 
ViT~\cite{dosovitskiy2021} introduced for computer vision tasks, where the images are split into patches and these are taken as the input 
sequence to a Transformer. In the same way, starting from a spin configuration $\sigma = (\sigma_1,\dots, \sigma_L)$, we split it into $N$ patches 
of $b$ elements: $\boldsymbol{x}_i = (\sigma_{(i-1)b + 1},\dots, \sigma_{(i-1)b + b})$, for $i = 1,\dots, N$ (the total number of sites must be
a multiple of $b$). The sequence of these patches is then used to compute the attention vectors. Then, a simplification of the original ViT is 
considered, taking the attention weights only depending on positions $i$ and $j$, but not on the actual values of the spins in these patches, 
thus leading to:
\begin{equation}\label{eq:attention_vec_pos}
\boldsymbol{A}^{\mu}_i = \sum_{j = 1}^N \alpha^{\mu}_{ij}\text{V}^{\mu}\boldsymbol{x}_j,
\end{equation}
where $V^{\mu}$ is a $r \times b$ matrix with $r=d/h$, and $d$ is the so-called {\it embedding dimension} that must be a multiple of the number 
of heads $h$. This approach is dictated by the fact that the attention weights should mainly depend on the relative positions among groups of 
spins and not on the actual values of the spins in the patches. This is expected to be true when the patches are far apart and is extended for 
generic positions $i$ and $j$. Finally, after the concatenation of the heads, a further linear projection is taken, before the non linearity, here 
chosen as \text{log[cosh($\cdot$)]}. This block can be repeated $n_l$ times before applying the output layer in which all the values are summed to 
obtain the logarithm of the ViT wave function $\Psi_{\text{ViT}}(\sigma)$ (see Fig.~\ref{fig:transf}). 

In order to study frustrated quantum spin models with a non-positive ground state (in the computational basis), we choose all the parameters to 
be {\it complex numbers}. Furthermore, a translationally-invariant wave function with $k=0$ can be easily defined by considering the following two
steps. First, we adapt the {\it relative positional encoding}~\cite{shaw2018} to periodic systems, taking $\alpha_{i,j}^{\mu}=\alpha_{i-j}^{\mu}$;
as a result, the number of variational parameters for computing the attention vectors~\eqref{eq:attention_vec_pos} is reduced from $O(L^2)$ to
$O(L)$. This procedure induces translational invariance between patches. To include also the one within patches, we perform the linear combination:
\begin{equation}\label{eq:psi_symm}
\tilde{\Psi}_{\text{ViT}}(\sigma) = \sum_{r = 0}^{b-1}\Psi_{\text{ViT}}(\text{T}_{r}\sigma),
\end{equation}
where $\text{T}_{r}$ is the translation operator. We emphasize that this approach requires a small summation (of $b$ terms), which does not grow
with the system size $L$.

The optimization process of all the complex parameters is obtained by using standard variational Monte Carlo techniques, namely the so-called 
Stochastic Reconfiguration approach (see the Supplemental Material \footnote{see Supplemental Material for details concerning the optimization of the neural network and the Monte Carlo sampling, which includes Refs.~\cite{sorella2005,becca2017}} for more details). In the following, we mainly take $n_l=1$, 
which represents the simplest possible adaptation of the Transformer architecture; indeed, even within this drastic assumption, we obtain excellent 
results in both gapless and gapped phases. At the end, we show the effect of a deeper network with $n_l>1$. All the simulations are performed by fixing 
the patch size $b=4$.

%%%%%%%%%%%%%%%%%%%%%%%%%%%%%%%%%%%%%%%%%%%%%%%%%%%%%%%%%%%%%%%%%%%%%%%%%%%%%%%%%%%%%%%%%%%%%%%%%%%%%%%%%%%%%%%%%%%%%%%%%%%%%%%%%
\begin{figure}
\includegraphics[width=\columnwidth]{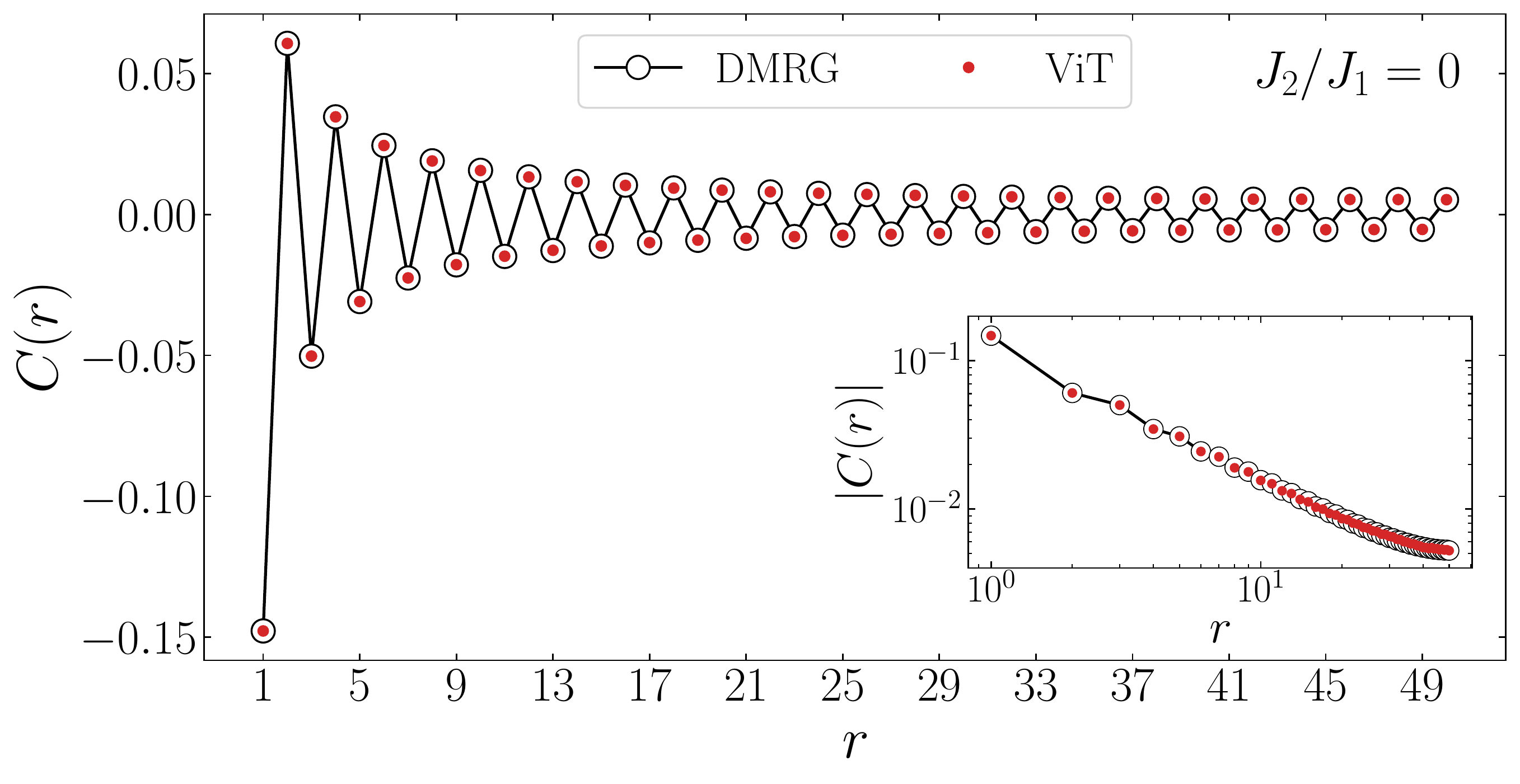}
\caption{\label{fig:J2=00_spinspin}
The isotropic spin-spin correlations in real space $C(r)$ as computed by the ViT wave function (full dots) for the unfrustrated Heisenberg model 
($J_2/J_1=0$) on a cluster with $L=100$ sites. The DMRG results are also shown for comparison (empty circles). Inset: Log-log plot of the same 
correlation function.}
\end{figure}

%%%%%%%%%%%%%%%%%%%%%%%%%%%%%%%%%%%%%%%%%%%%%%%%%%%%%%%%%%%%%%%%%%%%%%%%%%%%%%%%%%%%%%%%%%%%%%%%%%%%%%%%%%%%%%%%%%%%%%%%%%%%%%%%%

{\it Results.}
We start by discussing how the accuracy of the ViT wave function with one layer can be systematically improved by varying its two hyper-parameters, 
i.e., the number of heads $h$ and the ratio $r=d/h$. We consider a cluster with $L=100$ sites and three different values of the frustration ratio: 
$J_2/J_1=0$ (unfrustrated, gapless), $0.4$ (weakly-frustrated, gapped), and $0.8$ (strongly-frustrated, gapped); the reference energy 
is computed by using the standard DMRG approach (imposing periodic-boundary conditions on the Hamiltonian~\cite{pippan2010}). In Fig.~\ref{fig:acc}, 
we show the accuracy of the ground-state energy for the unfrustrated case as a function of $d/h$ fixing the number of heads $h$, and for the three 
values of $J_2/J_1$ when increasing the number of heads $h$, at fixed ratio $d/h$. Even though there is a general difficulty in reconstructing the 
exact sign structure in highly-frustrated regimes~\cite{viteritti2022,westerhout2020,szabo2020,park2021,bukov2021}, we obtain an excellent 
approximation of the correct energy for all the values of $J_2/J_1$ that have been considered, e.g., an accuracy $\Delta \varepsilon \lesssim 0.1 \%$
for $J_2/J_1=0.8$ and $\Delta \varepsilon \approx 0.01 \%$ for $J_2/J_1=0.4$.

Let us now move to the analysis of the correlation functions.
From the previous results, we choose $h=8$ and $d/h=1$ as a good compromise between accuracy and complexity, for which the network can be trained 
on $L=100$ sites in a few hours on ten CPUs or in a few minutes on a GPU. The spin-spin correlations are defined as
\begin{equation}\label{eq:spinspin}
C^{\nu\nu}(r) = \frac{1}{L}\sum_{R = 0}^{L-1}\langle \hat{S}^{\nu}_{R}\hat{S}^{\nu}_{R + r}\rangle,
\end{equation}
where $\nu=x$, $y$, or $z$ and $\langle \dots \rangle$ represents the expectation value over the variational quantum state. In particular,
we focus on isotropic spin-spin correlations $C(r) = [C^{zz}(r)+C^{xx}(r)+C^{yy}(r)]/3$ and the corresponding structure factor in Fourier space
$S(k) = \frac{1}{L}\sum_{r = 0}^{L - 1}\text{e}^{ikr}C(r)$. In Fig.~\ref{fig:J2=00_spinspin}, we show the results of the real-space correlations 
$C(r)$ for the unfrustrated Heisenberg model ($J_2/J_1 = 0$) on a cluster with $L=100$ sites, comparing them to the DMRG outcomes (with 
periodic-boundary conditions). Remarkably, the ViT {\it Ansatz} is able to match the DMRG calculations at all distances, demonstrating that the 
global structure of the multi-head attention layer is able to build the algebraic long-range tail.

\begin{figure}
\includegraphics[scale=0.315]{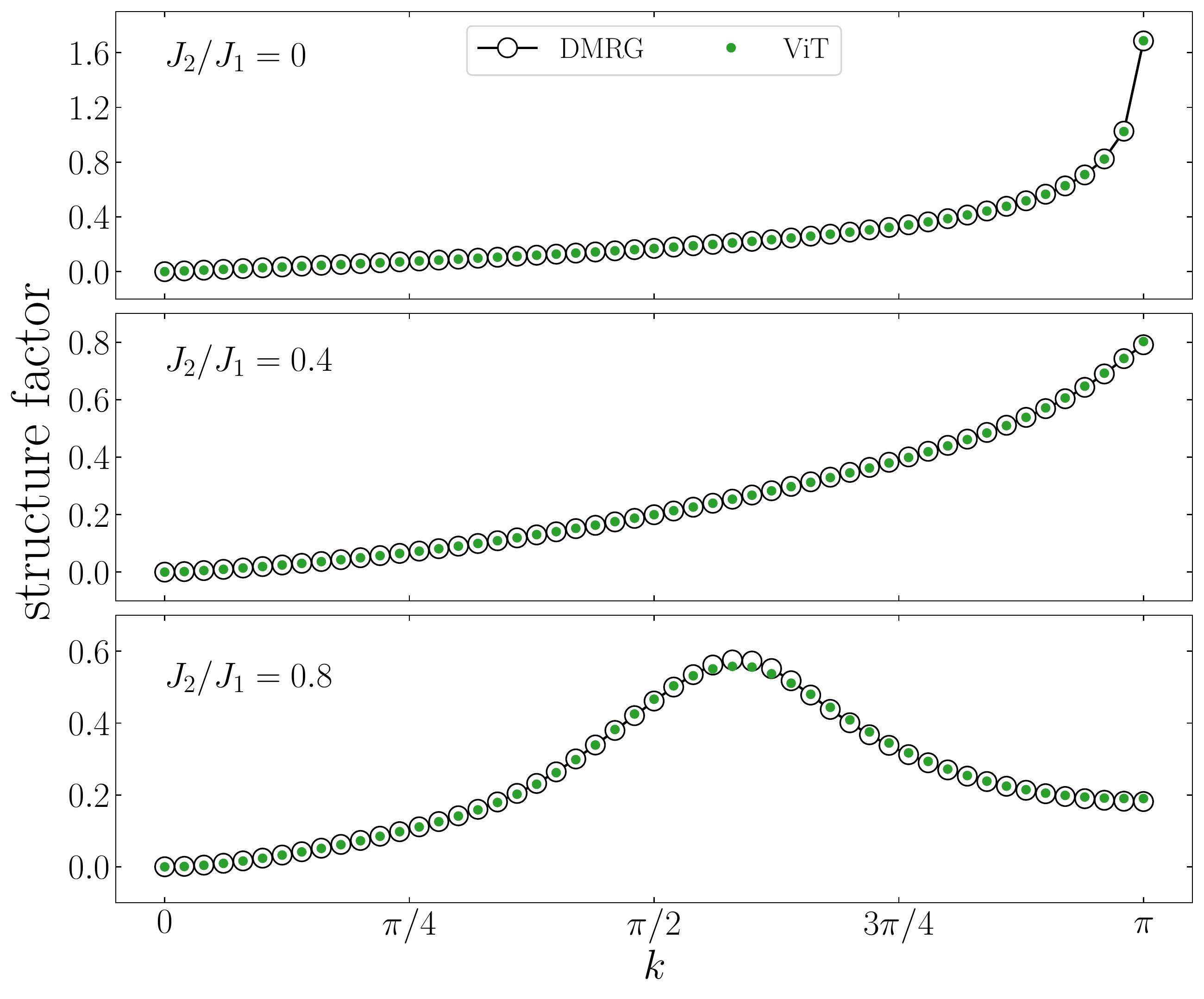}
\caption{\label{fig:Sk}
The spin-spin structure factor $S(k)$ as computed by the ViT wave function (full dots) for $J_2/J_1=0$ (upper panel), $J_2/J_1=0.4$ (middle panel) 
and $J_2/J_1=0.8$ (lower panel) on a cluster with $L=100$ sites. The DMRG results are also shown for comparison (empty circles).}
\end{figure}

The high flexibility of the ViT state is also demonstrated by considering the three different regimes, with commensurate (i.e., $S(k)$ peaked at $k=\pi$) or incommensurate (i.e., $S(k)$ peaked at $k \neq\pi$) 
correlations, see Fig.~\ref{fig:Sk}.

%%%%%%%%%%%%%%%%%%%%%%%%%%%%%%%%%%%%%%%%%%%%%%%%%%%%%%%%%%%%%%%%%%%%%%%%%%%%%%%%%%%%%%%%%%%%%%%%%%%%%%%%%%%%%%%%%%%%%%%%%%%%%%%%%

\begin{figure}
\includegraphics[width=\columnwidth]{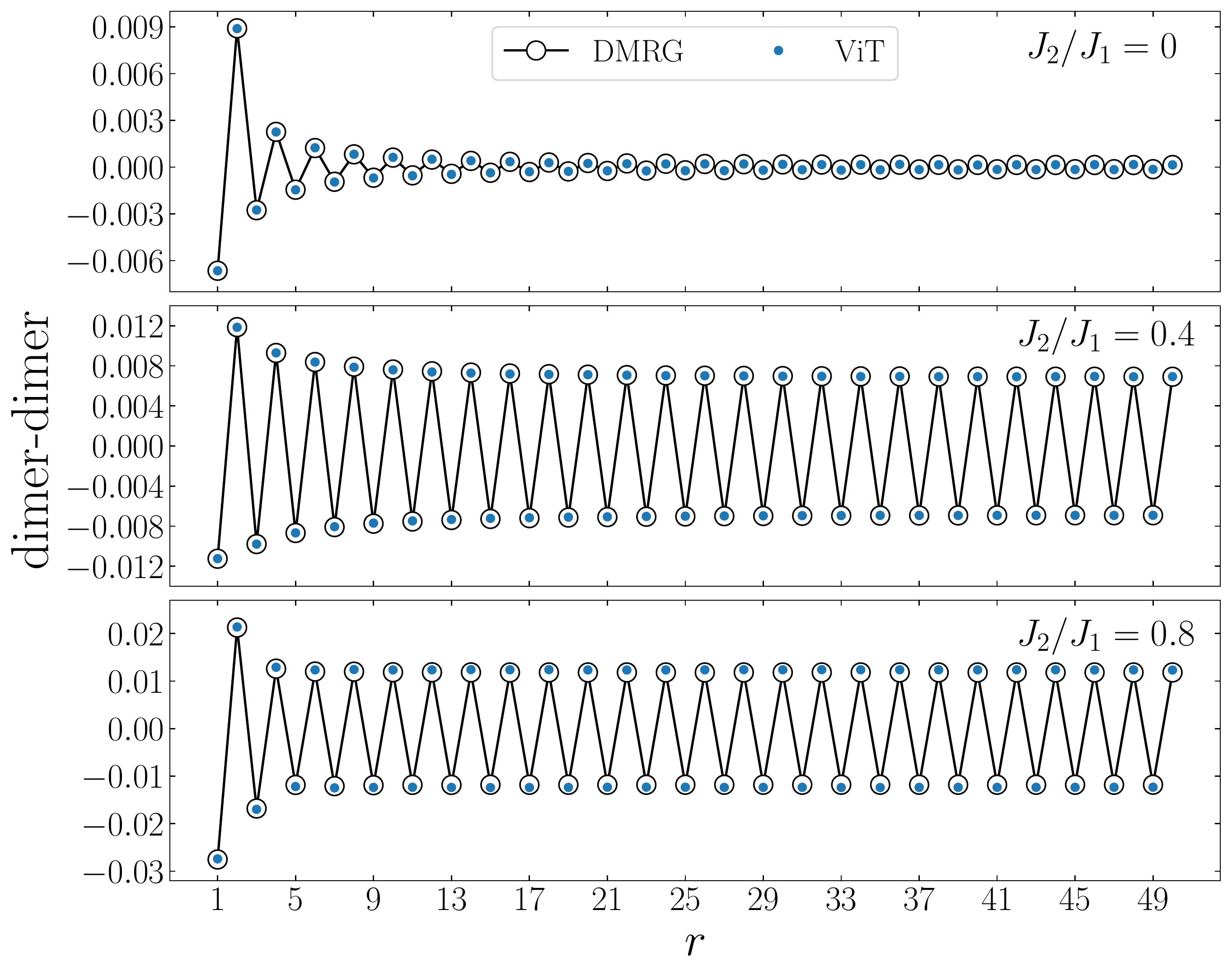}
\caption{\label{fig:Dimer}
Dimer-dimer correlations as computed by the ViT wave function (full circles) for $J_2/J_1=0$ (upper panel), $J_2/J_1=0.4$ (middle panel) and 
$J_2/J_1=0.8$ (lower panel) on a cluster with $L=100$ sites. The DMRG results are also shown for comparison (empty circles).}
\end{figure}
%%%%%%%%%%%%%%%%%%%%%%%%%%%%%%%%%%%%%%%%%%%%%%%%%%%%%%%%%%%%%%%%%%%%%%%%%%%%%%%%%%%%%%%%%%%%%%%%%%%%%%%%%%%%%%%%%%%%%%%%%%%%%%%%%

The gapped phase is characterized by a finite dimer order (implied by the two-fold degeneracy of the ground state, in the thermodynamic limit). 
On any finite system, there is an exponentially small gap between the two states, with $k=0$ and $k=\pi$, and the insurgence can be detected 
from the connected dimer-dimer correlations:
\begin{equation}\label{eq:dimerdimer}
D(r) = \frac{1}{L}\sum_{R=0}^{L-1}\langle \hat{S}^z_{R}\hat{S}^z_{R+1}\hat{S}^z_{R+r}\hat{S}^z_{R+r+1}\rangle - [C^{zz}(r=1)]^2,
\end{equation}
where $C^{zz}(r=1)$ is the $z$ component of the spin-spin correlation function at distance $r=1$ defined in eq.~\eqref{eq:spinspin}. Notice 
that this definition considers only the $z$ component of the spin operators~\cite{capriotti2003}. 
In Fig.~\ref{fig:Dimer}, we show the results for the three values of $J_2/J_1$ considered in this work. Again, the agreement with DMRG 
calculations is excellent in all cases, and the ViT state is able to perfectly reproduce the presence of dimer order.

\begin{figure}
\includegraphics[width=\columnwidth]{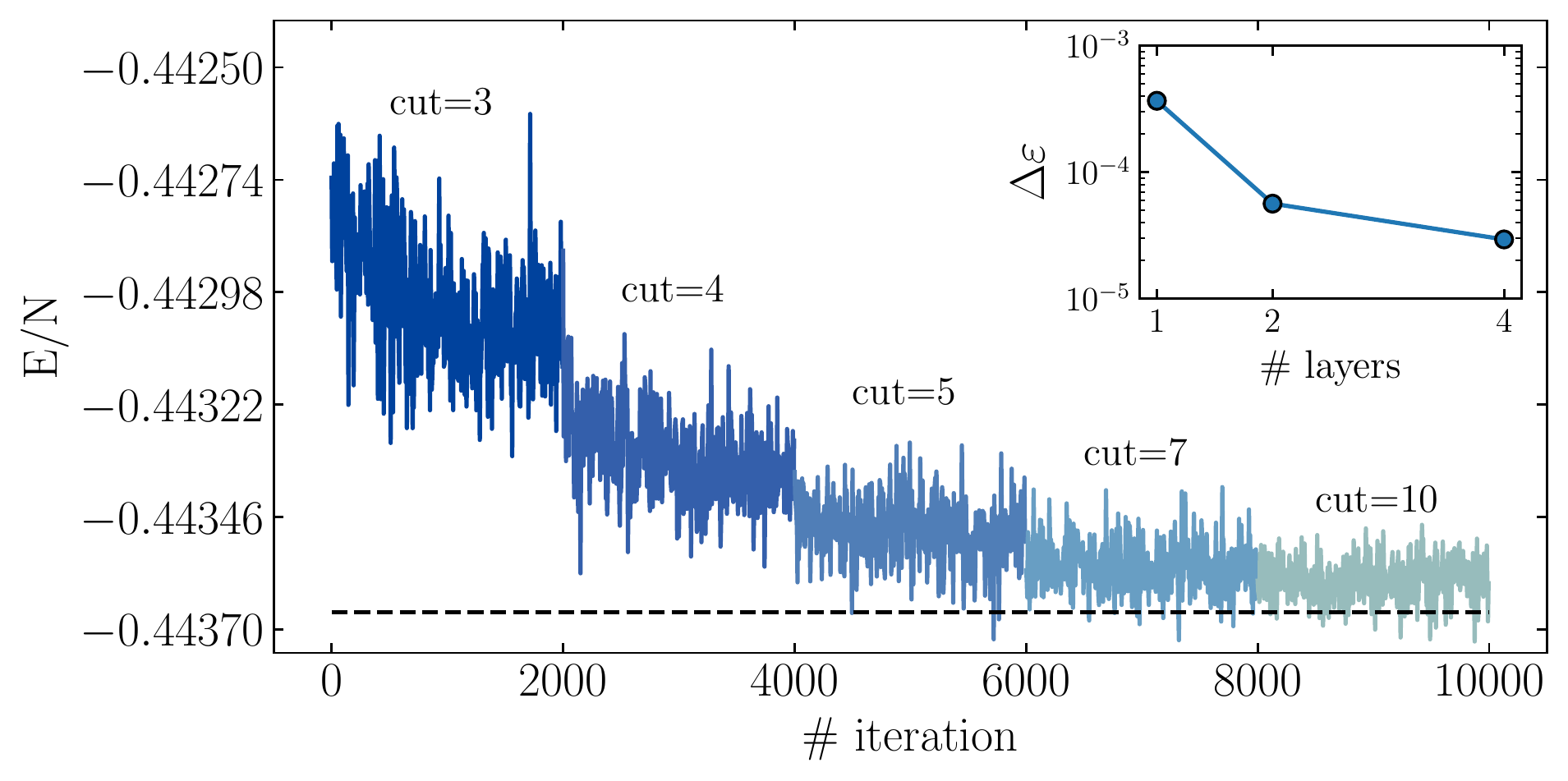}
\caption{\label{fig:optimization} 
Optimization of the DeepViT with $n_l=4$, where each layer has $h=2$ and $d/h=2$, for the Heisenberg model with $L=40$ sites. Along the process,
the cut in the attention is fixed and progressively increased from $1$ to $10$ (the first two values are not shown for better readability). At 
the end, once the cut has been completely relaxed, the full translational invariance is restored [see Eq.~\eqref{eq:psi_symm}] to compute the 
accuracy in the energy. Inset: Relative error $\Delta\varepsilon$ of the DeepViT wave function by varying the number of layers. The reference 
energy is computed by DMRG~\cite{pippan2010} with a bond dimension up to $\chi=600$ obtaining $E/J_1=-0.443663$.}
\end{figure}

{\it DeepViT.}
The ViT wave function can be systematically improved by stacking multiple Transformer layers, i.e., $n_l>1$. The optimization of deep networks is difficult with standard protocols, then we develop a procedure based on the physical interpretation of the attention weights. We start by setting for each head and layer $\alpha_{i-j} = 0$ if $|i-j| > {\rm cut}$, with ${\rm cut} < L/b$, training only the 
remaining weights. Small cut values (e.g., ${\rm cut}=1$) are good starting points for stable optimizations. Then the cut is relaxed until
reaching $L/b$, where all-to-all connections among the inputs of each layer are restored. As an example, the results for the Heisenberg model with 
$L=40$ are shown in Fig.~\ref{fig:optimization}. Here, we take $n_l=4$ (each layer has $h=2$ and $d/h=2$) and perform the optimization stages with
${\rm cut}=1,\dots,10$. Every time, when the cut is relaxed, the accuracy of the energy improves. We stress that the optimization is performed 
without Marshall sign prior.

%%%%%%%%%%%%%%%%%%%%%%%%%%%%%%%%%%%%%%%%%%%%%%%%%%%%%%%%%%%%%%%%%%%%%%%%%%%%%%%%%%%%%%%%%%%%%%%%%%%%%%%%%%%%%%%%%%%%%%%%%%%%%%%%%
%%%%%%%%%%%%%%%%%%%%%%%%%%%%%%%%%%%%%%%%%%%%%%%%%%%%%%%%%%%%%%%%%%%%%%%%%%%%%%%%%%%%%%%%%%%%%%%%%%%%%%%%%%%%%%%%%%%%%%%%%%%%%%%%%

{\it Conclusions.}
We have introduced a promising class of variational wave functions, which are based upon Transformer neural-network architectures (in particular,
Vision Transformers). Their main advantages, with respect to previously defined {\it Ans\"atze}, is the mixing of {\it local} and {\it global}
structures, which makes them very flexible to describe a variety of different quantum phases, with both gapped and gapless spectra.  Remarkably,
even working with a relatively simple architecture, with $n_l=1$, excellent results are obtained for a frustrated spin model in one spatial
dimension. Generalizations to one-dimensional
models with long-range interactions (e.g., the Haldane-Shastry model~\cite{haldane1988,shastry1988}) or two-dimensional models, where ground-state 
properties are still under debate, are desirable and represent the topic for future investigations, including the calculation of long-range 
entanglement properties~\cite{zhang2011}. We expect that for these systems the depth of the network could be important to achieve competitive results with respect to state-of-art numerical methods.

\begin{acknowledgments}
We thank A. Laio and S. Goldt for having drawn our attention to Transformers and E. Tirrito for useful discussions about DMRG implementations, which 
have been performed within the iTensor library~\cite{itensor}. The variational quantum Monte Carlo and the ViT architecture were implemented in 
JAX~\cite{jax2018github}.
\end{acknowledgments}

\bibliography{references.bib}

\clearpage
\onecolumngrid

\section*{Supplemental Material}
\subsection*{Optimization of the variational wave function}
The goal of the optimization procedure is the minimization of the variational energy
\begin{equation}
    E(\mathcal{W}) = \langle \hat{H} \rangle_{\mathcal{W}} = \frac{\langle{\Psi_{\mathcal{W}}}|\hat{H}|{\Psi_{\mathcal{W}}}\rangle}{\langle{\Psi_{\mathcal{W}}}|{\Psi_{\mathcal{W}}}\rangle} \ ,
\end{equation}
with respect to the variational parameters $\{\mathcal{W}\}$. To perform the optimization we employ the \textit{Stochastic Reconfiguration} method \cite{sorella2005} which we briefly describe in the following (for a detailed description see reference \cite{becca2017}).\\
For each parameter $w_{\alpha} \in \{\mathcal{W}\}$ we define the corresponding operator $\hat{\mathcal{O}}_{\alpha}$, diagonal in the computational basis $\langle \sigma |\hat{\mathcal{O}}_{\alpha}| \sigma'\rangle = {\mathcal{O}}_{\alpha}(\sigma)\delta_{\sigma,\sigma'}$, whose matrix elements are
\begin{equation}
    {\mathcal{O}}_{\alpha} = \frac{\partial \log\left(\Psi_{\mathcal{W}}(\sigma)\right)}{\partial w_{\alpha}} \ .
\end{equation}
At each optimization step the variational parameters are updated according to
\begin{equation}\label{eq:update_par}
    w_{\alpha}' = w_{\alpha} + \eta\sum_{\beta}S_{\alpha,\beta}^{-1}f_{\beta} \ ,
\end{equation}
where $\eta$ is the \textit{learning rate}, an hyperparameter of the optimization process, $f_{\beta}$ are the \textit{forces}
\begin{equation}
    f_{\beta} = -\frac{\partial \langle \hat{H} \rangle_{\mathcal{W}}}{\partial w_{\beta}} = -2\Re\{ \langle \hat{H}^*\hat{O}_{\beta}\rangle_{\mathcal{W}} - \langle\hat{H}^*\rangle_{\mathcal{W}}\langle\hat{O}_{\beta}\rangle_{\mathcal{W}}\}  \ ,
\end{equation}
and $S^{-1}$ is the inverse of the \textit{covariance matrix}
\begin{equation}\label{eq:S_matrix}
    S_{\alpha, \beta} = \Re\{\langle \hat{O}^{\dagger}_{\alpha}\hat{O}_{\beta}\rangle_{\mathcal{W}} - \langle\hat{O}_{\alpha}^{\dagger}\rangle_{\mathcal{W}}\langle\hat{O}_{\beta}\rangle_{\mathcal{W}} \} \ .
\end{equation}

The expectation values $\langle \cdot\cdot\cdot \rangle_{\mathcal{W}}$, defined with respect to the probability distribution $|\Psi_{\mathcal{W}}(\sigma)|^2$, are estimated stochastically using the Metropolis algorithm with a sample size of $O(10^3 \div 10^4)$. In addition, given the SU(2) spin symmetry of the $J_1$-$J_2$ Heisenberg model, the sampling procedure for the study of the ground state properties can be limited to the $S^z=\sum_R S^z_R =0$ sector, thereby in order to conserve the total magnetization, nearest- and next-nearest neighbor spin exchanges are considered.
The convergence of the optimization process, for a chain of $L=100$ sites considered in this work, is achieved in $O(10^2\div 10^3)$ steps setting the learning rate $\eta$ to $10^{-2}\div 10^{-3}$. We point out that the $S$-matrix defined in Eq.~\eqref{eq:S_matrix} can be not invertible due to a redundant parametrization of the variational state, which usually happens when the wave function has a large number of parameters. In order to prevent numerical instabilities in the inversion of the matrix, we regularize it by shifting the diagonal elements of the $S$-matrix by a small perturbation $S_{\alpha, \alpha} \rightarrow S_{\alpha, \alpha} + \varepsilon$, typical values are $\varepsilon \sim 10^{-3}\div 10^{-4}$.\\

\end{document}